\documentclass[sigconf]{acmart}

\usepackage{multirow}
\usepackage{xcolor}
\usepackage{tabularray}
\usepackage{hyperref}
\usepackage{subcaption}

\usepackage{tabularx}

\usepackage{framed}
\usepackage[most]{tcolorbox}
\newtcolorbox{bluebox}[1][]{
  colback=blue!5!white,
  colframe=blue!75!black,
  boxsep=2pt,      % inner padding between frame and content
  top=2pt,         % space between frame and top content
  bottom=2pt,      % space between frame and bottom content
  left=5pt,        % space between frame and left content
  right=5pt,       % space between frame and right content
  width=\linewidth % set the width of the box to match the column width
}
\newtcolorbox{boxK}{
    colframe = cs-2!80!white,
    sharpish corners, % better drop shadow
    boxrule = 0pt,
    boxsep = 0pt,
    toptitle = 2pt,
    bottomtitle = 1pt,
    top = 4pt,
    bottom = 5pt,
    left = 5pt,
    right = 5pt,
    enhanced,
    fuzzy shadow = {0pt}{-1pt}{-0.5pt}{0.5pt}{black!20}, % {xshift}{yshift}{offset}{step}{options}
    fonttitle=\small\bfseries,
    fontupper=\small,
    title=Main Findings%Key Takeaways
}
\usepackage{blindtext}
\usepackage{balance}
\balance

\AtBeginDocument{%
  }

\begin{document}

\title[ACE: Automated Technical Debt Remediation]{ACE: Automated Technical Debt Remediation with Validated Large Language Model Refactorings}

\author{Adam Tornhill}
\orcid{XXX}
\affiliation{%
  \institution{CodeScene}
  \city{Malmö}
  \country{Sweden}
}
\email{adam.tornhill@codescene.com}

\author{Markus Borg}
\orcid{XXX}
\affiliation{%
  \institution{CodeScene and Lund University}
  \city{Malmö}
  \country{Sweden}
}
\email{markus.borg@codescene.com}

\author{Nadim Hagatulah}
\orcid{XXX}
\affiliation{%
  \institution{Lund University}
  \city{Lund}
  \country{Sweden}
}
\email{nadim.hagatulah@cs.lth.se}

\author{Emma Söderberg}
\orcid{XXX}
\affiliation{%
  \institution{Lund University}
  \city{Lund}
  \country{Sweden}
}
\email{emma.soderberg@cs.lth.se}

\renewcommand{\shortauthors}{Tornhill et al.}

\begin{abstract}
The remarkable advances in AI and Large Language Models (LLMs) have enabled machines to write code, accelerating the growth of software systems. However, the bottleneck in software development is not writing code but understanding it; program understanding is the dominant activity, consuming approximately 70\% of developers' time. This implies that improving existing code to make it easier to understand has a high payoff and -- in the age of AI-assisted coding -- is an essential activity to ensure that a limited pool of developers can keep up with ever-growing codebases.

This paper introduces Augmented Code Engineering (ACE), a tool that automates code improvements using validated LLM output. Developed through a data-driven approach, ACE provides reliable refactoring suggestions by considering both objective code quality improvements and program correctness. Early feedback from users suggests that AI-enabled refactoring helps mitigate code-level technical debt that otherwise rarely gets acted upon.
\end{abstract}




\maketitle

\section{Introduction} \label{sec:intro}
High-quality code is a competitive business advantage, enabling companies to ship features faster and with fewer defects. Yet, this advantage remains largely untapped as Technical Debt (TD) and poor code continue to consume up to 42\% of developers’ time~\cite{stripe_developer_2018}. Despite these alarming numbers, research suggests that only 7\% of organizations systematically track TD~\cite{martini_technical_2018}. Instead, software organizations routinely prioritize new feature development over refactoring~\cite{ampatzoglou_financial_2015,ernst_technical_2021}.

This issue could intensify with the advent of AI-assisted development. On the one hand, early empirical studies show that AI assistants can increase developer speed. A pioneering paper by Peng \textit{et al.} claims that GitHub Copilot makes software developers 56\% faster~\cite{peng_impact_2023}. Similarly, a recent randomized control trial at Google reports a 21\% improvement in task completion speed when developers used AI-enhanced features for code~\cite{paradis_how_2024}. On the other hand, neither of these studies evaluated the impact on code quality --- which leaves us wondering whether a new kind of TD generator may soon be deployed at scale in industry.

Several studies warn about the risks of naïvely applying AI assistants in software development. For example, Yeti\c{s}tiren \textit{et al.} report that GitHub Copilot generates correct code only 31\% to 65\% of the time~\cite{yetistiren_assessing_2022}. Furthermore, an analysis by Harding and Kloster of 150 million changed lines of code reveals ``disconcerting trends for maintainability'' when adopting AI assistants for code~\cite{harding_coding_2024}. This implies that the software developers of tomorrow might face a growing mountain of code of unknown quality, much of it written by machines rather than humans.

The growth of AI-generated code of unknown quality has substantial implications for software maintenance, which typically accounts for over 90\% of a typical product’s lifecycle costs~\cite{dehaghani_which_2013}. Moreover, research shows that developers spend about 57-70\% of their time understanding existing code, i.e., program comprehension is the dominant programming activity~\cite{minelli_i_2015,xia_2018}. When the code is plagued by TD, the situation is exacerbated. Consequently, optimizing for software maintainability is crucial for any long-term software development project.

At its core, optimizing for software maintainability means refactoring code. Refactoring is the process of improving the design of existing code without changing its behavior~\cite{fowler_refactoring_2018}. This seemingly simple definition has two important implications for automated refactoring: 
\begin{itemize}
    \item A change is only a refactoring if it improves the design. However, ``improve'' has been largely subjective. To automate refactoring, we need an objective standard for assessing improvements.
    \item A change is not a refactoring if it fails to preserve the behavior of the original code, such as when it introduces a bug. To automate refactoring, we need confidence that the tool adheres to this requirement.
\end{itemize}

These two implications have guided our solution-oriented work. To address the first, we employ CodeHealth, a code-level quality metric shown to correlate strongly with defect density and development velocity~\cite{tornhill_code_2022,borg_increasing_2024}. By comparing the CodeHealth before and after refactoring, we can objectively assess whether the design improved or not.

The second refactoring implication is particularly concerning when using Large Language Models (LLM) to increase the level of automation. Given the tendency of LLMs to hallucinate and produce incorrect code~\cite{yetistiren_assessing_2022}, a reliable and trustworthy refactoring tool needs to complement an LLM with guardrails. That way, the results can be validated and incorrect refactoring attempts discarded --- this is in line with many implementations of automatic program repair~\cite{le_goues_automated_2019}.

The remainder of this tool paper is organized as follows. Section~\ref{sec:bg} describes design rationales for the ACE development and Section~\ref{sec:related} presents an overview of closely related work. Section~\ref{sec:overview} describes an overview of ACE, whereas Section~\ref{sec:details} describes inner workings such as prompts and validation. Finally, Section~\ref{sec:feedback} shares early user feedback before Section~\ref{sec:conc} concludes the paper.

\section{Background} \label{sec:bg}
TD was first introduced by Ward Cunningham in 1992 as a metaphor for ``not-quite-right code'' leading to extra work for the next developer. A few years later, the term became popular in the rapidly growing agile software development movement. Unfortunately, it was often broadly used to describe arbitrary quality issues. The research community became interested in the phenomenon with a first international workshop on TD in 2010 and several attempts at defining TD followed. A joint effort at a Dagstuhl seminar in Germany in 2016 resulted in a definition~\cite{dagstuhl} that has stood the test of time: ``In software-intensive systems, technical debt is a collection of design or implementation constructs that are expedient in the short term but set up a technical context that can make future changes more costly or impossible. Technical debt presents an actual or contingent liability whose impact is limited to internal system qualities, primarily maintainability and evolvability.''

Several researchers have surveyed tools for maintainability prediction and TD management over the last five years. In 2020, Lenarduzzi \textit{et al.} identified 60 maintainability prediction tools in a literature review~\cite{lenarduzzi_survey_2020}, but 35 were disregarded as outdated research prototypes. They concluded that the most widely adopted tools are commercial.

Focusing on TD measurement, a large number of researchers presented an overview and comparison of commercial tools in 2021~\cite{avgeriou_overview_2021}. The study included tools that calculate TD principal or TD interest. Nine tools were investigated, and the authors reported SonarQube as the most widely used tool based on online discussions. More recently, Lenarduzzi \textit{et al.} conducted a detailed comparison of six tools, including SonarQube, across 47 Java projects~\cite{lenarduzzi_critical_2023}. Their findings showed a low level of agreement between the tools and highlighted that most of the tools suffer from many false positives. 

CodeScene is another commercial tool supporting TD management, although it was not included in the above surveys. The tool uses a relatively small set of code smells (compared to the tools investigated by Lenarduzzi \textit{et al.}~\cite{lenarduzzi_critical_2023}) to compute its CodeHealth metric, which it uses for TD identification. Recently, Borg \textit{et al.} reported that CodeHealth produces substantially fewer false positives than SonarQube in Java projects~\cite{borg_ghost_2024}. They performed a competitive benchmarking study using the Maintainability Dataset created by Schnappinger \textit{et al.} as the ground truth~\cite{schnappinger_defining_2020}, which constitutes the most reliable dataset with human-annotated file-level maintainability assessments.

No matter what tool is used to identify refactoring targets, acting on the output has so far mostly been left to the organizations using the tool. The main barriers to action are urgency and skill. First, improving existing code always competes with the time allocated for building new features~\cite{ernst_technical_2021}. Second, refactoring is an acquired skill, and junior developers might lack the expertise to effectively improve the code~\cite{wang_empirical_2009}.

ACE was developed to tackle these barriers. By reliably automating code refactoring, ACE has the potential to amplify software organizations in several ways. Specifically, ACE can:
\begin{itemize}
    \item Reduce time spent on manual refactoring.
    \item Simplify TD remediation, freeing developers to focus on the creative work of adding new features.
    \item Elevate team skill levels by providing examples of effective code improvements.
\end{itemize}

As a baseline for ACE, we have presented a large-scale benchmarking study of popular LLMs on real-world refactoring tasks~\cite{tornhill_refactoring_2024}. The study involved more than 100,000 real-world CodeHealth issues in open-source codebases. Various LLMs were prompted to refactor the specific code smells, and the preservation of behavior was evaluated using the accompanying test suites. In the study, we also assessed if the CodeHealth indeed improved after the refactoring. 

The study revealed that the best-performing LLM generated functionally correct refactorings only 37\% of the time. This finding was corroborated in a recent paper on using LLMs for extract method refactorings by Pomian \textit{et al.}, which reported that LLMs provide valid and useful suggestions in 24\% of their attempts~\cite{pomian_next-generation_2024}. These results make it clear that we cannot use out-of-the-box AI models or tools that merely wrap an LLM API to reliably refactor code. Instead, a more promising approach is to use generative AI to create a pool of potential solutions and then apply validation guardrails.

ACE implements these guardrails by validating the AI-generated output and assigning a confidence level to the refactorings. By discarding incorrect solutions, 98\% of the remaining AI-generated refactorings improve CodeHealth while retaining the original behavior~\cite{tornhill_refactoring_2024}. That is, ACE elevates the precision from 37\% to 98\%. This precision, however, comes at a cost in terms of reduced recall. Out of the box, an LLM always provides output, resulting in 100\% recall. ACE reaches a recall of 52\%, thanks to its Contextual LLM Selection discussed in Section~\ref{sec:overview}. Thus, ACE enables confident refactoring of more than half of the detected code smells.

\section{Related Work on Refactoring} \label{sec:related}
Recent research in automated code refactoring explores a variety of approaches that use LLMs to improve code quality by removing code smells and reducing TD. Pomian \textit{et al.} introduce EM-Assist~\cite{pomian_next-generation_2024}, a refactoring tool that exploits LLMs to automatically suggest and perform extract method refactorings. They evaluated EM-Assist on 1,752 real-world extract method scenarios, finding that 76.3\% of the LLM-generated suggestions were hallucinations. To address this, the authors used static analysis techniques to validate simple program structures and program slicing to further improve the suggested candidates. While EM-Assist achieves higher recall than traditional tools such as JDeodorant and GEMS, it remains reactive, requiring developers to manually identify long methods for refactoring, and the tool is limited to Java and Kotlin, as a plugin for the IntelliJ IDEA.

In a broader empirical study, Liu \textit{et al.} showed that targeted prompting allows LLMs such as GPT-4 to detect 86.7\% of real-world refactoring opportunities across 20 different Java projects~\cite{liu_2024}. However, 7.7\% of the proposed refactorings introduced errors, necessitating the need for safeguards. The author proposes RefactoringMirror, which is a strategy that first identifies the refactoring change the LLM proposes by using a refactoring detection tool, e.g. ReExtractor, and then reapplies the suggested refactoring proposal details via a refactoring engine, such as the one provided by IntelliJ IDEA.

Another metric-driven refactoring tool is LiveRef~\cite{fernandes_2022}, that visually flags identified code smells with colors, through an aggregation of more than 20 software maintainability metrics, such as number of Lines of Code (LoC) and cyclomatic complexity. The color-coded severity indicator allows developers to prioritize refactoring opportunities. The authors' empirical evaluation showed that developers who used LiveRef applied refactorings more frequently and improved maintainability metrics faster than those who relied on manual methods. However, similar to EM-Assist, LiveRef is limited to Java and the IntelliJ IDEA as a plugin and has a long analysis latency on the order of seconds.

Shirafuji \textit{et al.} demonstrates a method of using an LLM, GPT-3.5-turbo, with few-shot prompting to refactor Python programs into simpler, more readable, and maintainable versions~\cite{shirafuji_2023}. Their method generated 10 candidate solutions per program, with 95.68\% yielding at least one valid solution, determined by unit tests to ensure functional correctness. The refactorings achieved a 17.35\% reduction in cyclomatic complexity and a 25.84\% decrease in LoC, showing that a few-shot prompting method can guide an LLM to produce refactorings that improve the original code. However, the study was limited to introductory Python problems and focused primarily on reducing complexity, even when the original code is already in a good state. Unlike ACE, which uses CodeScene to automatically identify methods with code smells as candidates for refactoring.

ACE differentiates itself by being a more self-contained tool while also expanding its applicability. Similarly to LiveRef, ACE uses a deterministic and metric-driven detection method, but is based on the CodeHealth metric, to identify code smells and refactoring opportunities with minimal latency. In contrast to RefactoringMirror, ACE applies mostly self-contained validations to LLM-generated refactoring suggestions to mitigate hallucinations without fully relying on external tooling. Furthermore, ACE increases the scope of automated code refactoring by supporting multiple programming languages and aiming to support multiple IDEs, which makes ACE a more robust tool for TD remediation.

\section{ACE Overview} \label{sec:overview}
In developing ACE, we leverage CodeScene output in two ways. First, we select a subset of the code smells that build up the CodeHealth score as refactoring targets. Second, we use CodeHealth as a fundamental part of the validation step described in Section~\ref{sec:validation}. As CodeScene is available as an IDE extension in VS Code, ACE follows suit and is currently available within the same ecosystem. At the time of writing, the release of ACE as a plugin for IntelliJ is imminent. A complete ACE use case is available as a screencast\footnote{\url{https://www.youtube.com/watch?v=7KQ1oXysFvc}}. 

\begin{figure*}
    \centering
    \includegraphics[width=\textwidth]{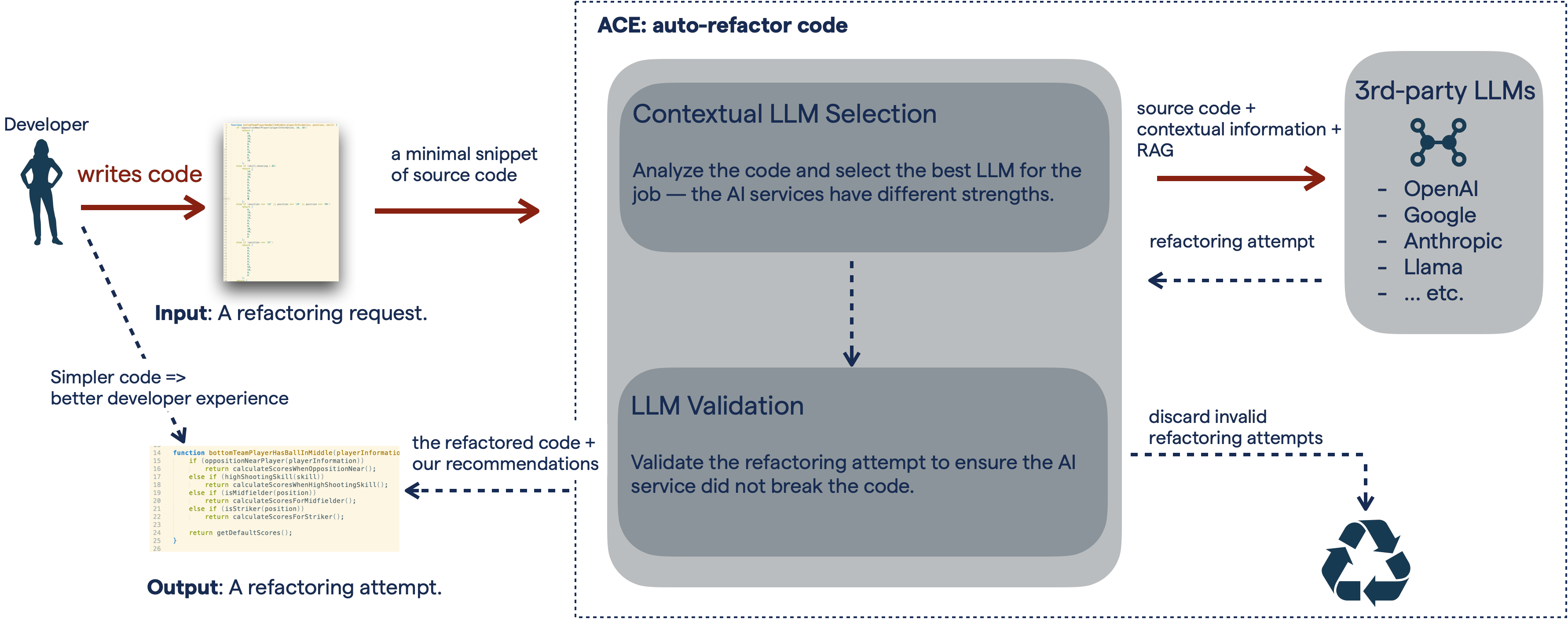}
    \caption{The high-level architecture of the ACE tool for reliable LLM-enabled refactoring.}
    \Description{This figure shows a high-level architecture of the ACE tool.}
    \label{fig:ace_architectural_overview}
\end{figure*}

Figure~\ref{fig:ace_architectural_overview} shows an overview of ACE. The back-end architecture, alongside a set of LLMs, is based on two key components:

\begin{enumerate}
  \item Contextual LLM Selection: The benchmarking study revealed that LLMs have different strengths and weaknesses~\cite{tornhill_refactoring_2024}. ACE takes advantage of this by using a machine learning model to select the most appropriate LLM based on the properties of the code to be refactored. The multi-LLM approach improves ACE's recall and decouples its application logic, facilitating the integration of new LLMs as they become available. The current version of ACE uses LLMs provided by OpenAI, Anthropic (Claude), Google, and the LLaMA family of models.
  \item LLM Validation: The responsibility of the validator is to ensure that the changed code is indeed a refactoring given the two implications presented in Section~\ref{sec:intro}. The validator does this by a combination of static code analysis techniques.
\end{enumerate}

\begin{figure*}
    \centering
    \includegraphics[width=\textwidth]{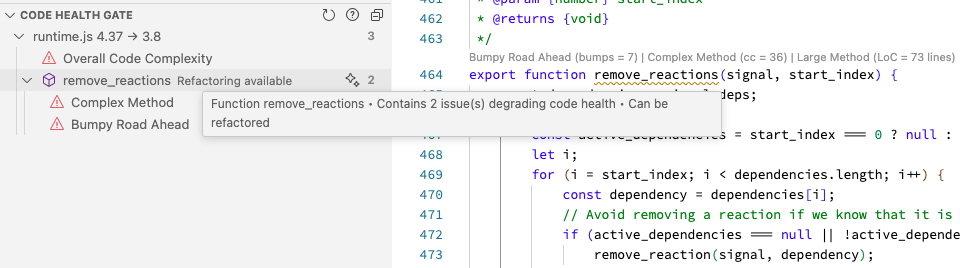}
    \caption{ACE offers automated refactoring of issues discovered in the CodeHealth quality gate.}
    \Description{This figure shows how ACE offers automated refactoring of issues discovered in the CodeHealth quality gate.}
    \label{fig:ace_quality_gate}
\end{figure*}

Developers interact with ACE through their IDE. There are two main interaction points for ACE, each one of them supporting a separate use case in the space of TD management:
\begin{enumerate}
  \item Refactor Declining CodeHealth: When a local quality gate detects a decline in CodeHealth, ACE offers refactoring suggestions as shown in Figure~\ref{fig:ace_quality_gate}.
  \item Proactive Code Improvement: Developers can proactively improve existing code on-demand as shown in Figure~\ref{fig:ace_high_conf_solution}.
\end{enumerate}

Most developers do not actively scan codebases for refactoring opportunities. Instead, a refactoring tool must address a specific need at the right time~\cite{murphy-hill_recommendation_2014}. For ACE, the local quality gate was a natural choice: it triggers when a developer lowers the CodeHealth. At that moment, the code change is fresh in the developer's mind, and offering automated refactoring has two advantages. First, developers might not know how to refactor the code effectively. Second, due to the pervasive time pressure in the software industry, code quality is often sacrificed for speed~\cite{li_systematic_2015} -- a micro-optimization that harms the macro progress. ACE overcomes these barriers, helping organizations manage TD by establishing a quality baseline for existing code, aligning with the ``Boy Scout Rule''~\cite{verdecchia2021building}. Via the quality gate, refactoring recommendations are delivered non-intrusively to avoid disrupting the developer's workflow.

Another common refactoring scenario arises when a developer wants to make a change but finds the existing code too complicated. Similarly, the team has decided to invest time into actively paying down their TD. In both situations, the developers look to elevate CodeHealth for a specific piece of code. ACE supports this use case by presenting code smells in the IDE and offering to fix them automatically.

Finally, it is worth emphasizing that ACE leaves the developer in control. While the refactorings are automated, the developer decides whether to accept them. To simplify the decision, ACE presents a summary of its review and the corresponding validation results. Even if a refactoring is not perfect, it can serve as a useful starting point, allowing developers to make minor tweaks and fill in the gaps themselves.

\section{ACE Architecture and Implementation} \label{sec:details}
This section describes key components in the ACE architecture and implementation. Furthermore, we explain which code smells are currently targeted.

\subsection{LLM Prompting Strategy}
ACE constructs prompts dynamically based on the specific code smell and its context. Input to this customization includes: i) programming language, allowing language-specific guidance (e.g., handling the "this" keyword in JavaScript), ii) source code of the function to be refactored, and iii) type of function, whether it is standalone or a class function, iv) code smell category, informing the refactoring strategy (e.g., applying extract method for complex conditionals), and v) code smell location within the function. These pieces of information are combined into a layered prompt that guides the LLMs to perform minimal and focused code changes to remove the code smell.

\subsection{LLM Validation Layer} \label{sec:validation}
\begin{figure}
    \centering
    \includegraphics[width=0.5\textwidth]{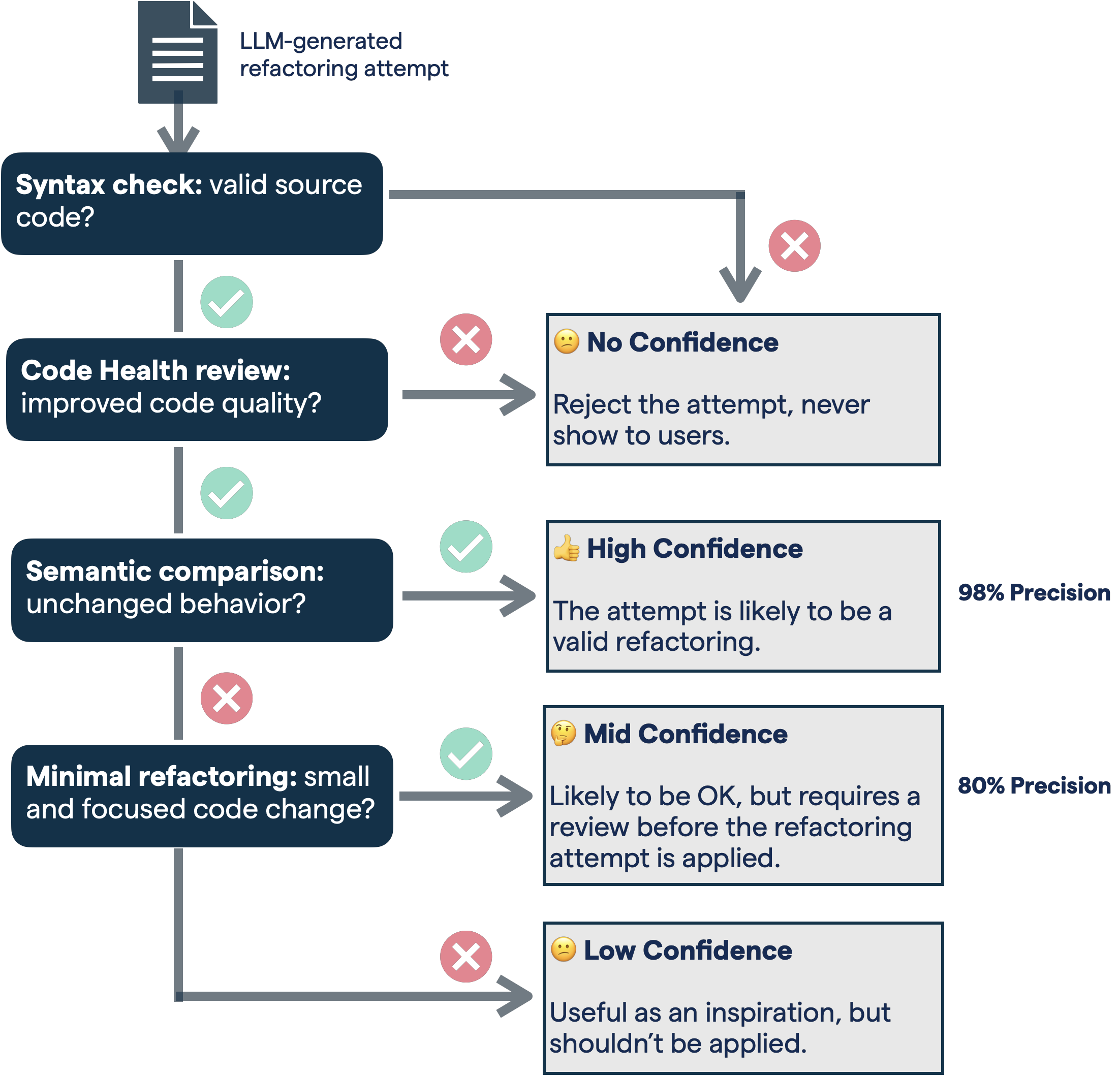}
    \caption{Flowchart for the validation layer for assigning a confidence level to each automated refactoring.}
    \Description{This figure shows a flowchart of the validation layer for assigning a confidence level to each automated refactoring.}
    \label{fig:ace_fact_checking}
\end{figure}

The dynamic prompting strategy and combining multiple AI services only explain parts of ACE's refactoring capabilities. The other fundamental component is the validation layer in ACE, whose high-level decision flow is shown in Figure~\ref{fig:ace_fact_checking}, which assigns a confidence level to each LLM-based refactoring. Note that ACE is a proprietary tool, and the exact details cannot be disclosed.

\begin{figure*}
    \centering
    \includegraphics[width=\textwidth]{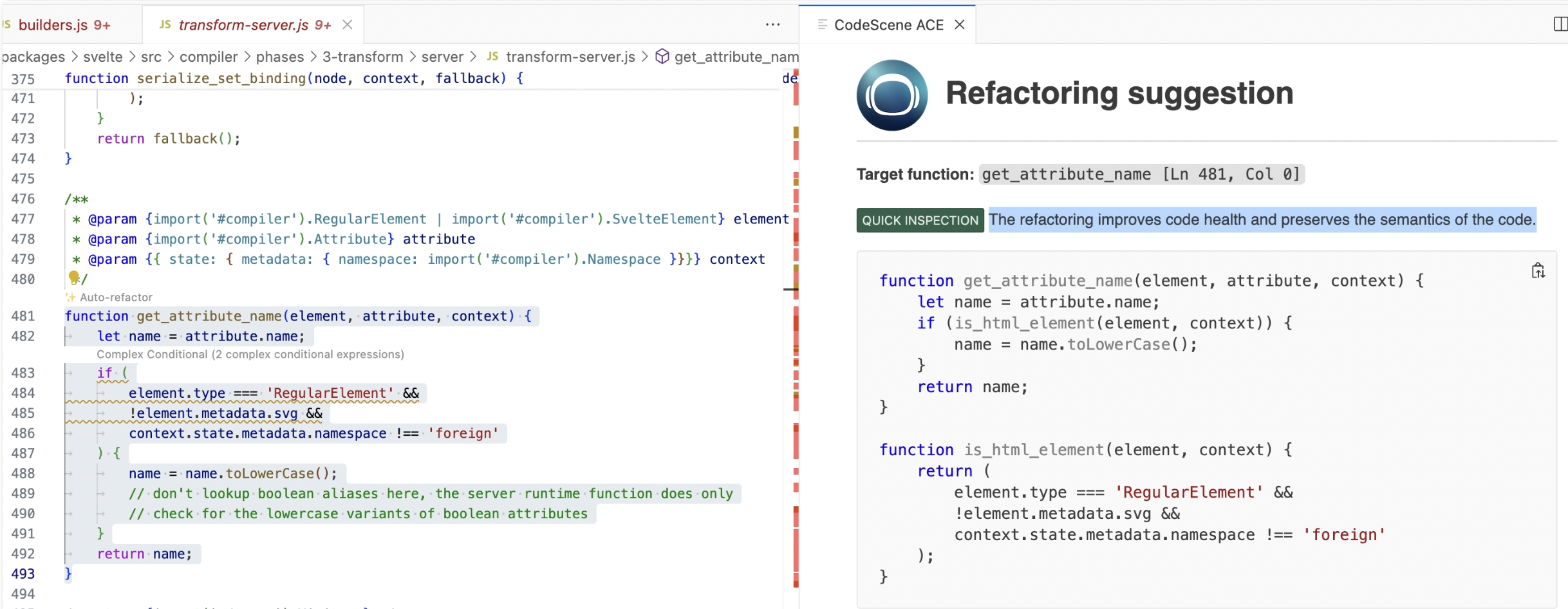}
    \caption{Example of a high confidence extract method refactoring.}
    \Description{This figure shows an example of a high confidence extract method refactoring.}
    \label{fig:ace_high_conf_solution}
\end{figure*}

\begin{figure*}
    \centering
    \includegraphics[width=\textwidth]{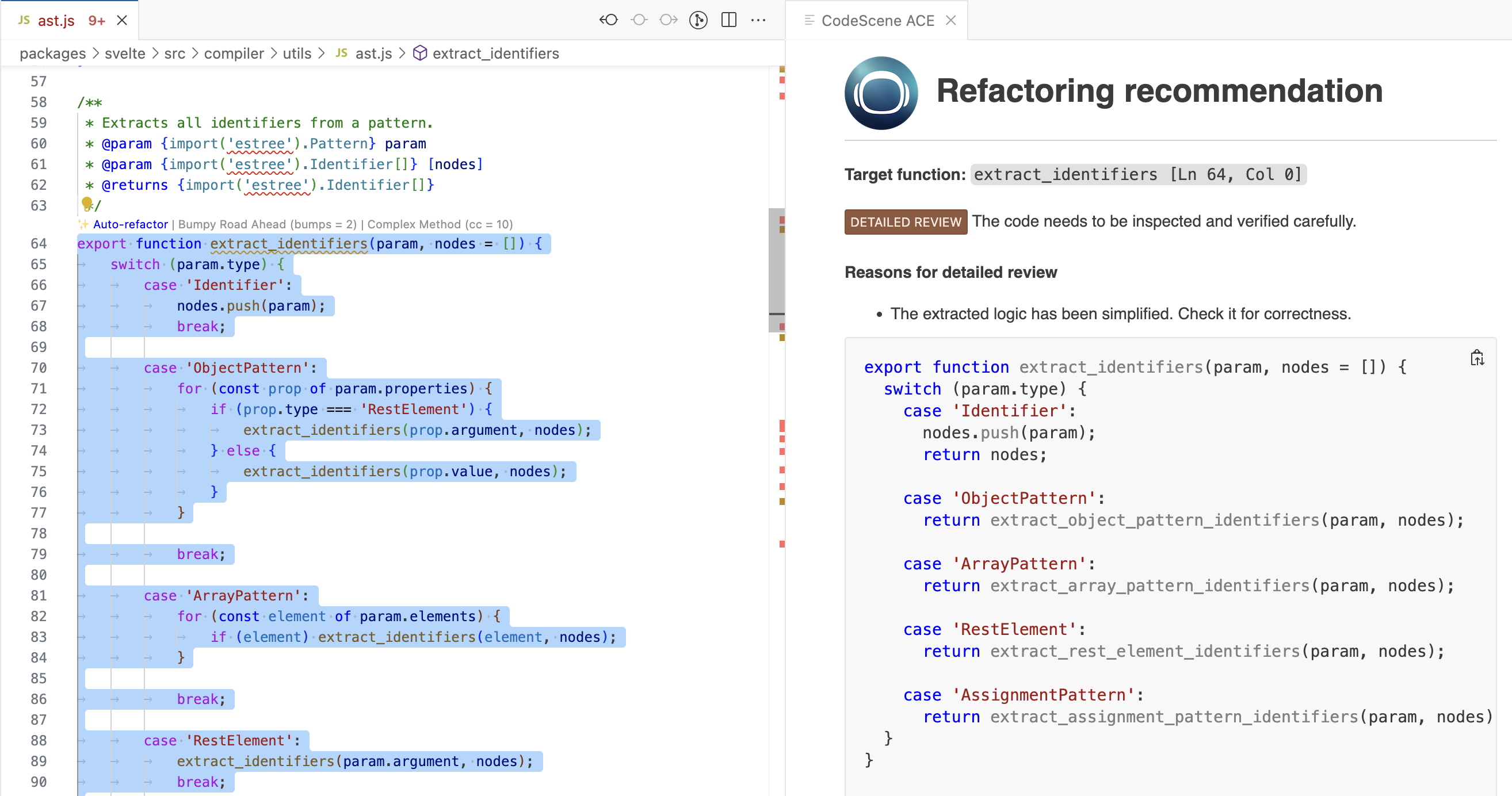}
    \caption{Example of a mid confidence refactoring that simplifies cyclomatic complexity.}
    %\caption{Example of straightening out a Bumpy Road in code while reducing the overall complexity of the function by simplifying its logic.}
    \Description{This figure shows an example of a mid confidence refactoring that simplifies cyclomatic complexity.}
    \label{fig:ace_svelte_example}
\end{figure*}

Low-confidence refactoring attempts are likely to break the code or make it worse. These are discarded by the ACE service and never shown to users. Mid- and high-confidence refactorings are presented to the user together with the rationale for the decisions. This rationale comes from the validation layer (see Figures~\ref{fig:ace_high_conf_solution} and~\ref{fig:ace_svelte_example}).

%Added paragraphs about syntatic and codehealth validation step. Too much information?
As shown in Figure~\ref{fig:ace_fact_checking}, there are multiple validation steps in ACE. The first step is syntactic validation, where a static analysis tool (e.g., ESLint for JavaScript) is used to verify that the LLM-generated refactoring output is syntactically valid. This step involves checking the original code and the refactored code for differences reported by the analysis tool. If the refactored code fails the check or introduces any additional warnings compared to the original code, then ACE discards the refactoring suggestion or lowers its confidence score. This step makes sure that the LLM-generated refactoring solution does not introduce any syntactic errors or warnings.

The second step validates that the targeted code smell has been resolved and that the overall CodeHealth has been improved by the LLM-generated refactoring suggestion. This step compares the code smell and CodeHealth score of the original and the refactored code for differences. If the targeted code smell remains or if the CodeHealth score is lower in the refactoring suggestion, then ACE discards that refactoring suggestion. If the code smell is removed and the CodeHealth score is improved, but a new, less severe code smell is introduced by the refactoring, then the confidence score will be lowered. This step ensures that the LLM-generated refactoring resolves the targeted code smell from the original code and increases the overall CodeHealth score without reducing the code quality.

The final steps in ACE's validation process involve semantic validation. These steps compare the original and refactored code to verify that the behavior is preserved and that only the expected minimal changes were introduced. During development, several recurring patterns were identified where AI-generated refactorings can break semantic correctness. These patterns are now captured and used to assess equivalence and assign a corresponding confidence level. If the semantic comparison confirms that the expected changes were made, a high confidence level is assigned. If minor but unintended changes are found, the result is marked with mid confidence. More substantial or unclear deviations result in low confidence. The expected changes are defined through a combination of code smell-specific validations and general language-specific rules, such as detecting empty method stubs or avoiding redundant re-implementations of existing logic.

It is important to note that assessing semantic equivalence is a largely unsolved research problem in the area of automated program repair~\cite{le-cong_invalidator_2023}. Consequently, ACE never aimed at solving semantic equivalence in general -- which is an undecidable problem according to Rice's theorem. Rather, the validation is specifically tailored to the set of code smells identified via the CodeHealth metric. This allows ACE to constrain the validation process to a finite set of structural changes that can be learned by our internal machine learning models. The quality of ACE's data lake, containing hundreds of thousands of refactorings with known ground truths, is the secret sauce, not the algorithms.

\subsection{Supported Code Smells and Programming Languages}
ACE is an evolving product. In its first release, the tool supports automated refactoring of five common code smells, all part of the larger aggregated CodeHealth:
\begin{itemize}
    \item Complex Conditional: expressions inside a branch (e.g., if, for, while) containing multiple, logical operators such as AND or OR.
    \item Complex Method: many conditional statements inside a function, measured using cyclomatic complexity.
    \item Deep Nested Logic: the nesting depth as measured by the number of if-statements and/or loops inside other program branches.
    \item Bumpy Road: a function with sequential chunks of Deep Nested Logic.
    \item Large Method: large functions as measured by the lines of code.
\end{itemize}

ACE currently supports automated refactoring of JavaScript, TypeScript, and Java. Our first priority has been on the dominant front-end languages, followed by Java. Next, we plan to extend ACE to cover additional back-end languages such as C\#, C, and C++.

\section{Early Feedback: Dogfooding and Limitations} \label{sec:feedback}
ACE was evaluated by two audiences to ensure a product-market fit. First, our development team used ACE internally. Although the precision and recall numbers were promising, subjective assessments from developers -- the target audience -- were crucial during development. Once the internal team found ACE valuable, the next step was to test it with external users in a controlled Alpha release.

One pilot user, an experienced tech lead, commented \textit{``I'm surprised with ACE's suggestions, as they are quite similar to what I would have come up with myself, saving me time and cognitive effort.''} Another software developer, a self-described ``AI skeptic,'' remarked \textit{``I like how ACE breaks down complex components into atomic parts with appropriate naming.''}

In fact, one of the key advantages of using LLMs for refactoring is their ability to name and decompose convoluted elements. Since approximately 70\% of code consists of identifiers, good naming is paramount for readability and maintainability~\cite{deissenboeck_concise_2006}.

The Alpha testing phase highlighted some early limitations that now inform the ACE roadmap:
\begin{itemize}
    \item \textbf{Applicability:} JavaScript and TypeScript are traditional front-end technologies. Much TD is on the back-end, necessitating support for more programming languages. This motivated us to implement support for Java, and more back-end languages are in the works.
    \item \textbf{LLM Scaling:} Our initial data showed a drop in LLM output quality for functions larger than 70 LoC, resulting in lower recall as function size increased. Our recall improvements stretched the maximum refactorable function length to 130 lines of code, and we aim to continue expanding this limit. 
    \item \textbf{Scope:} The first ACE version focuses on function-level code smells (e.g., Complex Conditional and Bumpy Road) to avoid breaking APIs. As a result, far from all code quality problems can currently by automatically fixed. Expanding the refactoring scope of ACE is an important direction for future work.
\end{itemize}

\section{Conclusion} \label{sec:conc}
ACE is a tool designed to automate refactoring and improve the design of existing code. By combining the creative capabilities of LLMs with a robust validation layer, ACE delivers high-confidence refactorings with 98\% precision, outperforming out-of-the-box LLMs.

Introducing an AI coding assistant that mitigates complex code smells safely allows users to optimize code for understanding -- the most human-intensive aspect of coding -- not just the occasional task of writing new code. An even more compelling potential of ACE is its promise of automated TD remediation. Every business manager is aware of TD, but few prioritize it -- and even fewer manage their debt actively~\cite{martini_technical_2018}.

ACE is a proprietary solution, but free to download and try. The tool currently supports automated refactoring for five code smells, with the capability to detect 25. Future versions of ACE will expand its refactoring scope. Similarly, an important future direction is to expand the number of supported programming languages to cover additional back-end languages like C\#, C++, and PHP.

\section*{Acknowledgments}
This work was partially supported by the Wallenberg AI, Autonomous Systems and Software Program (WASP) and partly by the Competence Centre NextG2Com funded by the VINNOVA program for Advanced Digitalisation with grant number 2023-00541.

\bibliographystyle{ACM-Reference-Format}
\bibliography{references}

\end{document}